\documentclass[prc,twocolumn,superscriptaddress,showpacs,preprintnumbers,amsmath,amssymb]{revtex4-1}
\usepackage{dcolumn}
\usepackage{bm}
\usepackage{color}
\usepackage{amssymb}
\usepackage{amsmath}
\usepackage{graphicx}
\usepackage{amsfonts}
\usepackage{slashed}
\usepackage{pstricks}
\usepackage{float}
\usepackage{hyperref}
\usepackage{array}
\allowdisplaybreaks
\usepackage{dcolumn}
\usepackage{epsf}
\usepackage{natbib}
\begin{document}
\title{Perturbation Theory of Nuclear Matter with a  Microscopic Effective Interaction}
\author{Omar Benhar}
\affiliation{INFN and Dipartimento di Fisica, ``Sapienza'' Universit\`a di Roma, I-00185 Roma, Italy} 
\author{Alessandro Lovato}
\affiliation{Physics Division, Argonne National Laboratory, Argonne, IL 60439}

\date{\today}

\begin{abstract}
An updated and improved version of the effective interaction based on the Argonne\textendash Urbana nuclear Hamiltonian\textemdash derived using the 
formalism of Correlated Basis Functions (CBF) and the cluster expansion technique\textemdash is employed to obtain a number of properties of 
cold nuclear matter at arbitrary neutron excess within the formalism of many-body perturbation theory. The numerical results\textemdash including 
the ground-state energy per nucleon, the symmetry energy, the pressure, the compressibility, and the single-particle spectrum\textemdash are discussed in the  
context of the available empirical information, obtained from measured nuclear properties and heavy-ion collisions.
\end{abstract} 

\index{}\maketitle

\section{Introduction}

Nuclear matter can be thought of as a giant nucleus consisting of $Z$ protons and $A-Z$
neutrons\textemdash in the  $A, Z~\to~\infty$ limit\textemdash  interacting through nuclear forces only. 
Besides being a necessary intermediate step towards the description of atomic nuclei, theoretical studies of such a system, which greatly 
benefit from the simplifications granted by translation invariance, provide the 
basis for the development of accurate models of matter in the neutron star interior. 

The ultimate goal of nuclear matter theory, clearly stated over forty years ago in the seminal paper of H.A.~Bethe~\cite{Bethe}, is the {\em ab initio} 
determination of its properties from a microscopic description of the underlying dynamics.
Unfortunately, however, the use of perturbation theory to achieve this objective is severely hampered by the very nature of strong interactions. The observation that 
the central charge-density of nuclei, extracted from the measured electron scattering cross sections, 
is nearly independent of the mass number, $A$, for $A \gtrsim 16$, is in fact a clear indication that nuclear forces are strongly repulsive at short 
range. As a consequence, the matrix elements of the nucleon-nucleon potential between eigenstates of the non interacting system 
turn out to be large, and can not be treated as perturbations. 

The two main avenues to overcome the above problem are based either on the replacement of the bare nucleon-nucleon potential with an effective interaction,
derived taking into account the contribution of ladder diagrams to all orders~\cite{Gmatrix_RMP,Bethe}, or on the use of a basis of {\em correlated} states,
embodying non-perturbative interaction effects~\cite{CLARK197989,PW_RMP}. In recent years, it has been suggested that effective interactions suitable for 
perturbative calculations can also be obtained combining potentials derived within chiral perturbation theory and renormalization group evolution to low momentum.
However, the applications of this approach appear to be  confined to a rather narrow density region~\cite{SRG1,SRG2,SRG3}.

In the early 2000s, the authors of Ref.~\cite{PhysRevC.67.035504,Cowell:2004ng} exploited the formalism based on 
correlated states to derive a well behaved effective interaction and consistent current operators\textemdash suitable to carry out perturbative calculations of
the nuclear matter response to weak interactions\textemdash from a microscopic nuclear Hamiltonian. 
In Refs.~\cite{PhysRevLett.99.232501,Lovato:2012ux,Lovato:2013dva}, 
this approach has been extended and improved to take into account the effects of three-nucleon forces, which are known to 
play an important role at supranuclear densities.  The resulting effective interactions have been 
used to perform calculations of  a variety of nuclear matter properties of astrophysical interest,  including the shear viscosity and thermal conductivity coefficients~\cite{PhysRevLett.99.232501,PhysRevC.81.024305}  and the neutrino mean free path~\cite{Lovato:2012ux,Lovato:2013dva}. 

The potential of the approach based on perturbation theory and effective interactions obtained from correlated functions  
has been recently confirmed by systematic studies of the Fermi hard sphere system~~\cite{mecca_1,mecca_2}.  
      
In this article, we report the results of perturbative nuclear matter calculations carried out using an improved effective interaction, 
allowing a consistent treatment of systems with arbitrary neutron excess. 
 
The main features of the nuclear Hamiltonian and the derivation of the effective interaction are outlined in Section~\ref{TF}, while Section~\ref{results}
is devoted to the discussion of numerical results, including the ground-state energy, the symmetry energy, the pressure, the compressibility and the proton and neutron spectra
and effective masses. Finally, in Section~\ref{summary}
we summarize our findings, and lay down the prospects for future applications of our approach.

\section{Theoretical framework}
\label{TF}

In this section, we discuss the phenomenological model of nuclear dynamics employed in our work, and describe the procedure leading to 
the determination of the effective interaction.
 
\subsection{The nuclear Hamiltonian}

Within non relativistic Nuclear Many-Body Theory (NMBT), atomic nuclei, as well as infinite nuclear matter,  are described
in terms of point-like nucleons of mass $m$, whose dynamics are dictated by the Hamiltonian
\begin{equation}
H=\sum_i -\frac{\nabla_{i}^2}{2m}+\sum_{i<j}v_{ij}+\sum_{i<j<k} V_{ijk}\, .
\end{equation}
The complexity of nuclear forces clearly manifests itself in the deuteron. The fact that a two-nucleon bound state is only observed with total spin and ispospin $S~=~1$ and $T~=~0$ signals a strong spin-isospin dependence of the interaction, while the non vanishing electric quadrupole moment reflects a non spherically-simmetric charge-density distribution, implying in turn the presence of non-central forces. 

The nucleon-nucleon (NN) potential $v_{ij}$ is modelled in such a way as to reproduce the 
measured properties of the two-nucleon system, in both bound and scattering
states, and reduces to the Yukawa one-pion-exchange potential at large distances. 

Coordinate-space NN potentials are usually written in the form
\begin{equation}
v_{ij}=\sum_{p} v^{p}(r_{ij}) O^{p}_{ij} \ , 
\label{eq:NN_1}
\end{equation}
where $r_{ij} = |{\bf r}_i - {\bf r}_j|$ is the distance between the interacting particles, and the sum includes up to eighteen terms. 
The most prominent contributions are  those associated with the operators 
\begin{align}
O^{p \leq 6}_{ij} = [1, (\bm{\sigma}_{i}\cdot\bm{\sigma}_{j}), S_{ij}]
\otimes[1,(\bm{\tau}_{i}\cdot\bm{\tau}_{j})]  \ ,
\label{av18:2}
\end{align}
where $\bm{\sigma}_{i}$ and $\bm{\tau}_{i}$ are Pauli matrices acting in spin and isospin space, respectively, while the operator 
\begin{align}
S_{ij}=\frac{3}{r_{ij}^2}
(\bm{\sigma}_{i}\cdot{\bf r}_{ij}) (\bm{\sigma}_{j}\cdot{\bf r}_{ij})
 - (\bm{\sigma}_{i}\cdot\bm{\sigma}_{j}) \ , 
 \label{S12}
\end{align}
reminiscent of the potential describing the interaction between two magnetic dipoles, accounts for the occurrence of non-spherically-symmetric forces. 

The potential models obtained including the six operators of Eqs.~\eqref{av18:2}-\eqref{S12} 
explain deuteron properties and the $S$-wave scattering phase shifts up to pion production threshold. In order to describe the $P$-wave,  one has to include two additional components involving  the momentum dependent operators
\begin{align}
O^{p=7,8}_{ij}= ( {\boldsymbol \ell}\cdot{\bf S} ) \otimes[1,(\bm{\tau}_{i}\cdot\bm{\tau}_{j})]  \ ,
\end{align}
where ${\boldsymbol \ell}$ denotes the angular momentum of  the relative motion of the interacting particles.

The operators corresponding to $p~=~7,\ldots,14$ are associated with the non-static components of the NN interaction, 
while those corresponding to $p=15,\ldots,18$ account for small violations of charge symmetry. All these terms are included in the state-of-the-art 
Argonne $v_{18}$ (AV18) potential~\cite{Wiringa:1994wb}, providing a fit of the scattering data 
collected in the Nijmegen database, the low-energy nucleon-nucleon scattering parameters and deuteron properties with a reduced chi-square $\chi^2\simeq 1$. 

The results reported in this article have been obtained using the so-called Argonne $v_{6}^\prime$ (AV6P) interaction, which is not simply a
truncated version of the full AV18 potential\textemdash obtained  neglecting the contributions with $p>6$ in Eq.~\eqref{eq:NN_1}\textemdash 
but rather its reprojection on the basis of the six spin-isospin operators of Eqs.~\eqref{av18:2}-\eqref{S12}~\cite{Wiringa:2002ja}.  

The inclusion of the additional three-nucleon (NNN) term, $V_{ijk}$, is needed to explain the binding energies of the three-nucleon systems and the 
saturation properties of isospin-symmetric nuclear matter (SNM).  
The derivation  of  $V_{ijk}$  was first discussed in the pioneering work of  Fujita and Miyazawa~\cite{Fujita:1957zz}. They argued that its main 
component originates from two-pion-exchange processes in which a NN interaction leads to the excitation of one of the participating nucleons
to a $\Delta$  resonance, which then decays in the  aftermath  of the interaction with a third nucleon. Commonly used phenomenological
models of the NNN force, such as the Urbana IX (UIX) potential adopted in this work~\cite{Pudliner:1995wk}, are written in the form 
\begin{align}
V_{ijk}=V_{ijk}^{2\pi}+V_{ijk}^{N} \ ,
\end{align}
where $V_{ijk}^{2\pi}$ is the attractive Fujita-Miyazawa term, while $V_{ijk}^{N}$ is a purely phenomenological repulsive term.
The parameters entering the definition of the above potential are adjusted in such a way as to reproduce the ground state energy of
the three-nucleon systems and the equilibrium density of SNM, when used in conjunction with the AV18 NN interaction.

It has to be emphasized that within the framework of NMBT the determination of the nuclear Hamiltonian implies minimum theoretical bias, because the two- and three-nucleon systems are solved exactly, and the equilibrium properties of SNM can be computed with great accuracy.

As a final remark, we note that  local NN potentials derived within the alternate framework of chiral perturbation theory are also written as in Eq.~\eqref{eq:NN_1}
~\cite{Gezerlis:2013,piarulli:2015}.  Because local versions of the chiral NNN potentials~\cite{Lynn:2016} have the
same spin-isospin structure of the UIX force, the scheme described in this paper can be readily applied using chiral nuclear Hamiltonians. 

\subsection{The CBF effective interaction}

The formalism of Correlated Basis Functions (CBF) is
based on the variational approach to the many-body problem with strong forces, first proposed by R.~Jastrow back in the 1950s \cite{jastrow}.
Within this scheme,  the trial ground state of the nuclear hamiltonian is written in the form
\begin{equation}
\label{def:corrfun}
|\Psi_0 \rangle \equiv \frac{\mathcal{F}|\Phi_0\rangle}{\langle \Phi_0 | \mathcal{F}^\dagger  \mathcal{F} |\Phi_0\rangle^{{1/2}}} \ ,
\end{equation}
where $|\Phi_0\rangle$ is a Slater determinant built from single particle states $| \phi_\alpha \rangle$, with $\{ \alpha \}$ being the set of quantum numbers of
the states belonging to the Fermi sea. In the case of uniform matter  
at density $\rho = \nu k_F^3 /(6 \pi^2)$\textemdash where $k_F$ and $\nu$ denote the Fermi momentum and the degeneracy of momentum eigenstates, respectively\textemdash 
$| \phi_\alpha \rangle$ consists of a plane wave, with momentum ${\bf k}_\alpha $ such that $|{\bf k}_\alpha| \leq k_F$,  and the Pauli spinors associated with spin and isospin degrees of freedom.

The operator $\mathcal{F}$, describing the effects of correlations among the nucleons, is written as a product of two-body operators, whose structure mirrors the one of the AV6P potential. The resulting expression is
\begin{equation}
\mathcal{F}  \equiv \mathcal{S} \prod_{i<j} F_{ij}   \ ,
\end{equation}
with 
\begin{equation}
F_{ij}=\sum_{p=1}^6 f^p(r_{ij}) O^{p}_{ij}  \ .
\end{equation}
Note that the symmetrization operator $\mathcal{S}$ is needed to fulfill the requirement of antisymmetry of  $|\Psi_0\rangle$ under particle exchange,
since, in general, $[O^p_{ij},O^q_{jk}]\neq 0$.

The radial  dependence of the correlation functions $f^p(r_{ij})$ is determined from 
functional minimization of the expectation value of the Hamiltonian in the correlated ground state
\begin{equation}
\label{EV}
E_V = \langle \Psi_0 | H | \Psi_0 \rangle \ .
\end{equation}
The short-distance behavior is largely shaped by the strongly repulsive core of the NN potential, resulting in a drastic suppression of the probability to find two 
nucleons at relative distance  $r_{ij} \lesssim~1$~fm, while at longer distance the non-central, or tensor, components of interaction become prominent. 

The calculation of the variational energy of Eq.~\eqref{EV} involves severe difficulties. It can be efficiently carried out 
expanding the right-hand side in a series, whose terms describe the contributions of subsystems, or
clusters, involving an increasing number of correlated particles~\cite{CLARK197989}. The terms of the cluster expansion are represented by diagrams, that  
can be classified according to their topological structures.  Selected classes of diagrams can then be summed up to all orders solving a set of coupled 
non-linear integral equations\textemdash referred to as Fermi Hyper-Netted Chain/Single-Operator Chain (FHNC/SOC) equations~\cite{Fantoni:1974jv,Pandharipande:1979bv}\textemdash to obtain an accurate estimate of the ground state energy.

Accurate calculations of the expectation value of the nuclear Hamiltonian in the correlated ground state have been also 
carried out using the Variational Monte Carlo (VMC) method~\cite{Carlson:2014vla}.
Since VMC works in the complete spin-isospin space, which grows exponentially with $A$,  this approach is currently limited 
to nuclei with $A \leq 12$  by the available computational resources. However, the computational effort can be drastically 
reduced performing a cluster expansion similar to the one employed to derive the FHNC/SOC equations.  This scheme, known as Cluster Variational  Monte Carlo 
(CVMC)~\cite{pieper:1990,pieper:1992} has been recently exploited to calculate the ground-state properties of nuclei as large
as $^{16}$O and  $^{40}$Ca using  realistic phenomenological two- and three-nucleon 
potentials~\cite{Lonardoni:2017egu}.

Under the assumption that the correlation structure of the ground and excited states of the system be the same, the operator
 $\mathcal{F} $ obtained from the variational calculation of $E_V$ can be used to generate 
correlated excited states from Eq.~\eqref{def:corrfun} through the replacement  $|\Phi_0\rangle \to |\Phi_n\rangle$, with $|\Phi_n\rangle$
being any eigenstate of the non-interacting Fermi gas. 
The resulting correlated states span a complete, although non orthogonal, set, that can be used to carry out 
perturbative calculations within the scheme developed in Ref.~\cite{CBF1}. This approach, known as CBF perturbation theory, 
has been successfully applied to study a variety of fundamental nuclear matter properties, including the linear response functions~\cite{response,response2} and the two-point
Green's function~\cite{GF1,GF2}. 

In CBF perturbation theory, one has to evaluate matrix elements of the {\it bare} nuclear Hamiltonian, the effects of correlations being taken into account by the  transformation of the basis states describing the non interacting system. However, the same result can in principle be obtained transforming the  
Hamiltonian, and  using the Fermi gas basis. This procedure leads to the  appearance of an {\it effective} Hamiltonian suitable for use in 
standard perturbation theory, thus avoiding the non trivial difficulties arising from the use of a non-orthogonal basis~\cite{OCS}.

The CBF effective interaction is defined through the matrix element of the bare Hamiltonian in the correlated ground state, according to 
\begin{equation}
\label{def:veff}
\langle\Psi_0 | H | \Psi_0 \rangle = T_F + \langle \Phi_0 | \sum_{i<j} v_{ij}^\text{eff} | \Phi_0\rangle \ , 
\end{equation}
where $T_F$ denotes the energy of the non interacting Fermi gas, and the effective potential is written in terms of the 
same spin-isospin operators appearing in Eq.~(\ref{eq:NN_1}) as
\begin{equation}
v_{ij}^\text{eff}=\sum_{p} v^{\text{eff}\,, p}(r_{ij}) O^{p}_{ij} \ . 
\end{equation}
From the above equations, it is apparent that $v_{ij}^{\text{eff}}$ embodies the effect of correlations. As a consequence, it is well behaved at short distances, and can in principle be used to carry out perturbative calculations of any properties of nuclear matter. 

The authors of  Ref.~\cite{Cowell:2004ng} first proposed to obtain the effective interaction performing a cluster expansion of the left-hand side
of Eq.~\eqref{def:veff} and keeping the two-body cluster contribution only. While leading to a very simple and transparent expression
for $v_{ij}^{\text{eff}}$, however, this scheme was seriously limited by its inability to take into account the NNN potential $V_{ijk}$. In Ref.~\cite{PhysRevLett.99.232501} the effects 
of interactions involving more than two nucleons have been included through a density-dependent modification of the NN potential at intermediate range~\cite{LagPan}.

A groundbreaking improvement has been achieved by the authors of Refs.~\cite{Lovato:2012ux,Lovato:2013dva}, who explicitly took 
into account three-nucleon cluster contributions to the ground-state energy.
This procedure allows to describe the effects of three-nucleon interactions 
at fully microscopic level using the UIX potential. 

Note that the correlation functions $f^p(r_{ij})$ entering the definition of $v_{ij}^\text{eff}$ are not the same as those obtained from the minimization of 
the variational energy of Eq.~\eqref{EV}. 
They are adjusted so that the ground state energy computed at first order in $v_{ij}^{\text{eff}}$\textemdash that is, in the Hartree-Fock approximation\textemdash 
reproduces the value of $E_V$ resulting from the full FHNC/SOC calculation. 
In Refs.~\cite{Lovato:2012ux} and ~\cite{Lovato:2013dva}, this procedure
was applied, {\it independently}, to SNM and pure neutron matter (PNM). The effective interaction employed in this work, on the other hand, {\em simultaneously} describes
the density dependence of the energy per nucleon of {\it both} SNM and PNM. This feature is essential for astrophysical applications, because it allows  
to evaluate the  properties of nuclear matter at fixed baryon density and large neutron excess, which is believed to make up a large region of the neutron star interior.

\begin{figure}[h]
\includegraphics[width=7.5cm]{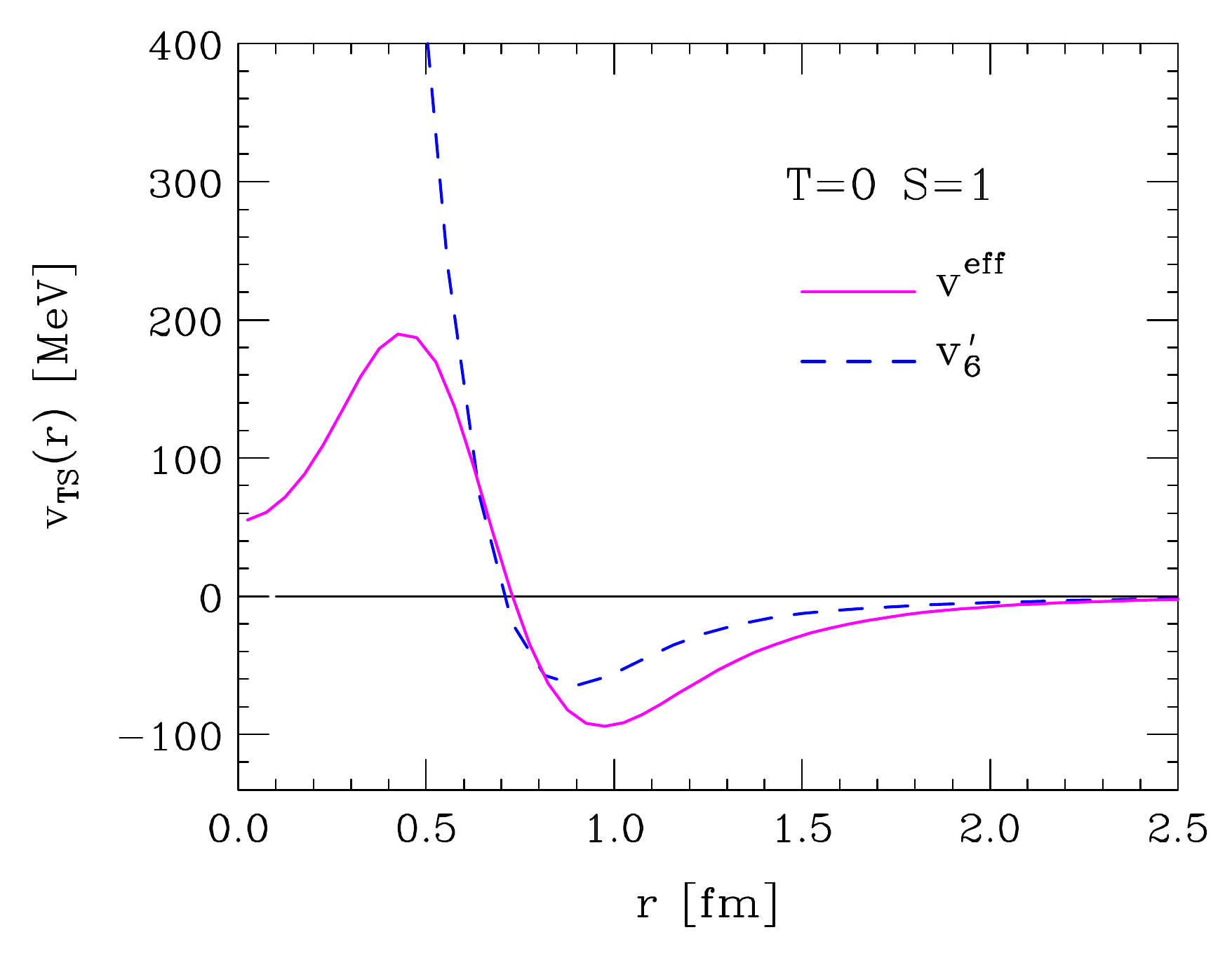}
\caption{Radial dependence of the  spherically-symmetric component of the bare AV6P potential (dashed line) and the CBF effective interaction (solid line)
in the spin-isospin channel corresponding to $S=1$ and $T=0$. The effective interaction has been computed setting $\rho=\rho_0$.\label{comp_v}}
\end{figure}

The radial dependence of the spherically-symmetric component of the potential describing the interaction of two nucleons coupled with total spin and isospin $S=1$ and $T=0$ 
is illustrated in Fig.~\ref{comp_v}. The solid and dashed lines correspond to the CBF effective interaction at $\rho = \rho_0$ and to the bare V6P potential, respectively. It clearly
appears that correlations significantly affect both the short- and intermediate-range behavior.

\section{Nuclear Matter Properties}
\label{results}

In the following, we will consider nuclear matter at baryon density 
\begin{equation}
\rho=\sum_\lambda \rho_\lambda=\rho \sum_\lambda x_\lambda \ , 
\label{eq:rho_asym}
\end{equation}
where $\lambda=1,2,3,4$ labels spin-up protons, spin-down protons, spin-up neutrons and spin-down neutrons,
respectively, the  corresponding densities being $\rho_\lambda=x_\lambda \rho$. In SNM  
 $x_1=x_2=x_3=x_4=1/4$,  while in PNM $x_1=x_2=0$ and $x_3=x_4=1/2$.

\subsection{Ground state energy}

At first order in the CBF effective interaction, the energy per baryon can be written in the form 
\begin{align}
\label{eq:E_HF}
\frac{E}{A}& = \frac{3}{5} \sum_{\lambda} x_\lambda \frac{k_{F,\lambda}^2}{2m}+\frac{\rho}{2}\sum_{\lambda\mu} x_\lambda x_\mu 
\int d^3 r \\ 
\nonumber
& \ \ \ \ \ \ \ \  \times \Big[v^\text{eff,d}_{\lambda\mu}({\bf r}) - v^\text{eff,e}_{\lambda\mu}({\bf r}) \ell(k_{F,\lambda} r)  \ell(k_{F,\mu} r)  \Big] \ , 
\end{align}
with the direct and exchange matrix elements of $v_{ij}^\text{eff}$ between spin-isospin states $| \lambda\mu\rangle$, given by
\begin{align}
\label{veffd}
v^\text{eff,d}_{\lambda\mu}({\bf r}_{ij}) & = \sum_{p} v^{p}(r_{ij}) \langle \lambda\mu | O^{p}_{ij} | \lambda\mu\rangle \ , \\
\label{veffe}
v^\text{eff,e}_{\lambda\mu}({\bf r}_{ij}) & = \sum_{p} v^{p}(r_{ij}) \langle \lambda\mu | O^{p}_{ij} | \mu\lambda\rangle \ . 
\end{align}
In Eq.~\eqref{eq:E_HF}, $k_{F,\lambda} = (6 \pi^2 \rho_\lambda)^{1/3}$ denotes the Fermi momentum of the particles of type $\lambda$, while
the function  $\ell(k_{F,\lambda} r)$,  referred to as Slater function, is trivially related to the density matrix in the absence of interactions, defined as
\begin{equation}
\label{slater}
\rho_\lambda \ell(k_{F,\lambda} r)\equiv  \frac{1}{V} \sum_\mathbf{k} e^{i\mathbf{k}\cdot \mathbf{r}} n_\lambda(k) \, ,
\end{equation}
where $n_\lambda(k) = \theta(k_{F,\lambda} - k)$ is the zero-temperature Fermi distribution and $V$ is the normalization volume. For the sake of completeness, the 
explicit expressions of the matrices $v^\text{eff,d}_{\lambda\mu}$ and  $v^\text{eff,e}_{\lambda\mu}$ are given in Appendix~\ref{matrices}.

\begin{figure}[h]
\includegraphics[width=8.5cm]{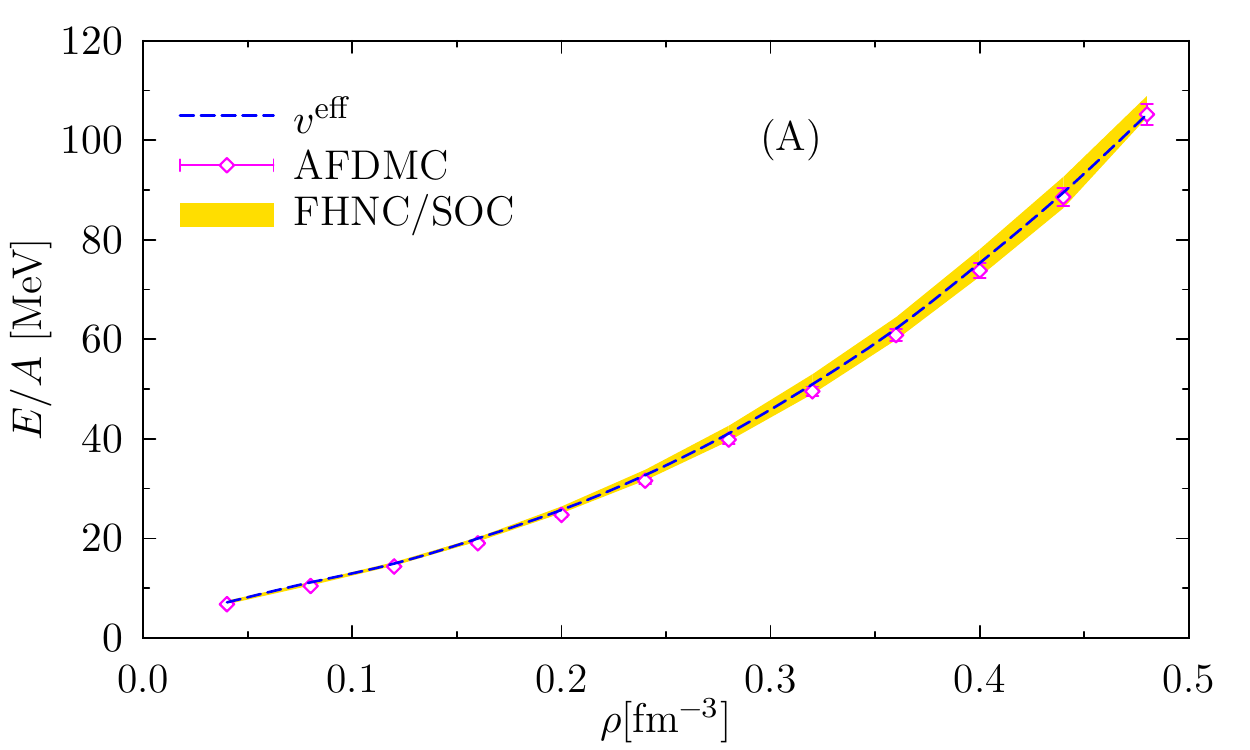} 
\includegraphics[width=8.50cm]{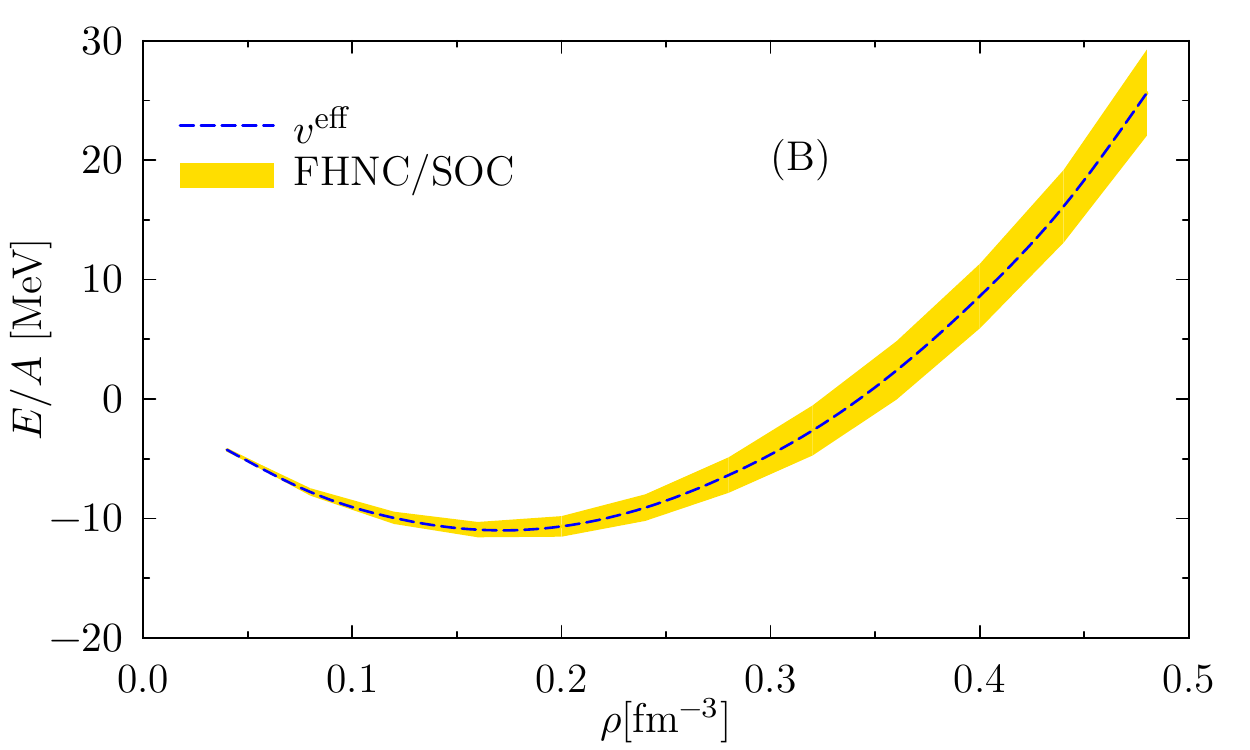}
\caption{Density dependence of the energy per nucleon of PNM (A) and SNM (B).  The solid lines show the results obtained using Eqs.~\eqref{eq:E_HF}-\eqref{slater} 
and the CBF effective interaction.The variational FHNC/SOC results are represented by the shaded regions, illustrating the uncertainty associated with the treatment of the kinetic energy~\cite{CLARK197989}, while the  open circles of panel (A) correspond to the PNM results obtained using the AFDMC technique. \label{fig:EoS}}
\end{figure}

\begin{figure*}[t]
\includegraphics[width=12.0cm]{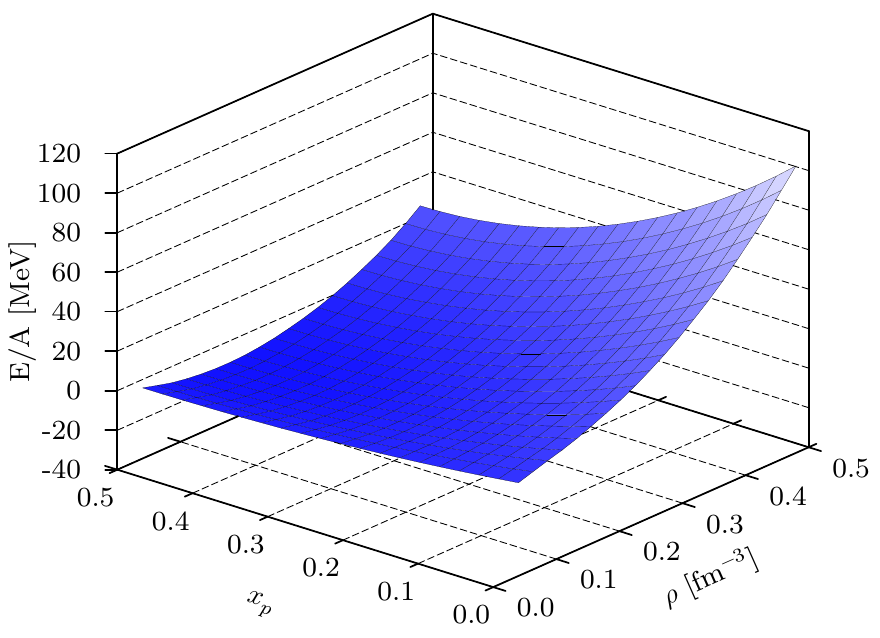}
\caption{Energy per nucleon of nuclear matter, computed as a function of baryon density and proton fraction using Eqs.~\eqref{eq:E_HF}-\eqref{slater} and the 
CBF effective interaction. \label{fig:EoS_asym}}
\end{figure*}

The solid lines of Fig.~\ref{fig:EoS} illustrate the density dependence of the energy per nucleon of PNM (A) and SNM (B), obtained from 
Eqs.~\eqref{eq:E_HF}-\eqref{slater}  with the CBF effective interaction. The shaded regions show the FHNC/SOC results obtained from the bare Hamiltonian, with the associated 
theoretical uncertainty arising from the treatment of the kinetic energy~\cite{CLARK197989}. For comparison, the results of a calculation carried 
out using the Auxiliary Field Diffusion Monte Carlo (AFDMC) technique~\cite{Schmidt199999} are also displayed. It clearly appears that the FHNC/SOC variational estimates, 
exploited as baseline for the determination of the CBF effective interaction, provide very accurate upper bounds to the ground state energy of PNM over the whole density range.
Note that the simplified AV6P~+~UIX Hamiltonian predicts the correct equilibrium density of SNM, $\rho_0 \approx 0.16 \ {\rm fm}^{-3}$, although the corresponding binding energy, $\sim 11$ MeV,  is below the empirical value of 16 MeV. However, it must be kept in mind that, because the kinetic and interaction energies largely cancel one another, 
a  $\sim 5$ MeV discrepancy in the ground-state energy translates into a $\sim 15\%$ underestimate of the interaction energy. This is consistent with the results of variational 
calculations of SNM performed with the full AV18+UIX Hamiltonian~\cite{akmal:1998}, yielding $E_0/A = -11.85\,{\rm MeV}$. The same Hamiltonian has been also found to underestimate
the binding energy of both $^{16}$O and $^{40}$Ca, by $2.83(3)\,{\rm MeV}/A$ and $3.63(10)\,{\rm MeV}/A$, respectively~\cite{Lonardoni:2017egu}.

Equations~\eqref{eq:E_HF}-\eqref{slater} have been also used to compute the energy per nucleon of unpolarized 
matter\textemdash corresponding to  $x_1=x_2$ and $x_3 = x_4$\textemdash   at fixed baryon density $\rho$ 
and proton density  $\rho_p = x_p \rho$, with $x_p=2x_1$, in the range $0 \leq x_p \leq 0.5$. The results of these calculations are displayed in Fig.~\ref{fig:EoS_asym}.

\subsection{Symmetry energy}

Consider again unpolarized matter with proton and neutron densities $\rho_p = x_p \rho$ and
$\rho_n = (1 - x_p)\rho$, respectively. The ground-state energy per nucleon can be expanded in series of powers of  
the quantity $\delta = 1 - 2 x_p = (\rho_n - \rho_p)/\rho$, providing a measure of neutron excess. The resulting expression reads (see, e.g., Ref.~\cite{marcello})
\begin{align}
\frac{1}{A} E_0(\rho, \delta) = \frac{1}{A} E_0(\rho, 0) + E_{\rm sym}(\rho) \delta^2 + O(\delta^4) \ ,
\label{quadratic}
\end{align}
where the symmetry energy 
\begin{align}
E_{\rm sym}(\rho)& = \left\{ \frac{ \partial^2 [E_0(\rho,\delta)/A] }{\partial \delta^2}\right\}_{\delta=0} \\
\nonumber & \approx  \frac{1}{A} E_0(\rho, 1) - \frac{1}{A} E_0(\rho, 0)
\end{align}
can be interpreted as the energy required to convert SNM into PNM.
The density dependence of  $E_{\rm sym}(\rho)$, that can be obtained expanding
around the equilibrium density of SNM, $\rho_0$, is conveniently characterized by the 
quantity
\begin{align}
\label{def:L}
L = 3 \rho_0 \left( \frac{ d E_{\rm sym}}{ d \rho} \right)_{\rho=\rho_0} \ .
\end{align}

Empirical information on $E_{\rm sym}(\rho_0)$ and $L$ have been extracted from 
data collected by laboratory experiments and astrophysical observations \cite{symmetry:expt}.  
The values resulting from our calculations, $E_{\rm sym}(\rho_0) = 30.9$ MeV and $L=67.9$ MeV, turn out to be compatible 
with those obtained from a survey of 28 analyses, carried out  by the authors of Ref.~\cite{symmetry:expt}, yielding  $E_{\rm sym}(\rho_0) = 31.6\pm2.66$ and
$L=58.9\pm16$ MeV. 

The density dependence of  the symmetry energy has been recently discussed in Ref.~\cite{russotto}, whose authors combined the 
results of isospin-dependent flow measurements carried out by the ASY-EOS Collaboration at GSI with those obtained from 
analyses of low-energy heavy-ion collisions~\cite{Tsang} and nuclear structure studies \cite{IAS,Brown,Zhang}. 

\begin{figure}[bth]
\includegraphics[width=8.0cm]{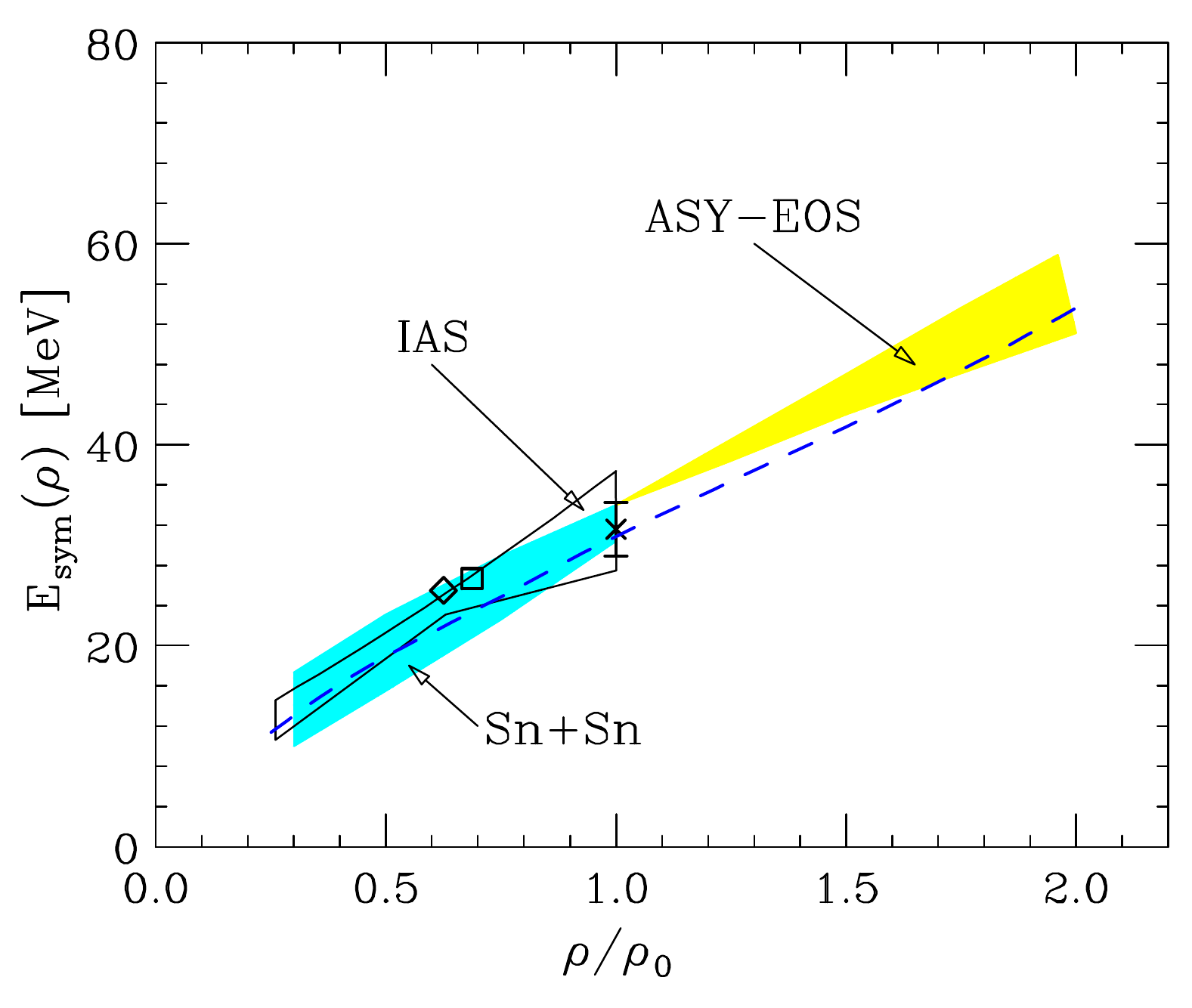}
\caption{ Density dependence of the symmetry energy of nuclear matter. The regions labelled ASY-EOS, Sn+Sn and IAS represent the results 
reported in Refs.\cite{russotto}, ~\cite{Tsang}, and \cite{IAS}, respectively, while the symbols correspond to the analyses of Refs.~ \cite{symmetry:expt} 
(cross with error bar), \cite{Brown} (diamond), and \cite{Zhang} (square). The results of the present work are displayed by the 
dashed line.
\label{E_sym}}
\end{figure}

Figure~\ref{E_sym} shows a comparison between  $E_{\rm sym}(\rho)$ resulting from our calculations and the empirical information 
reported in  Refs.\cite{symmetry:expt,russotto,Tsang,IAS,Brown,Zhang}. It is apparent that the theoretical results are compatible with experiments
at most densities.

\begin{figure}[h]
\includegraphics[width=7.5cm]{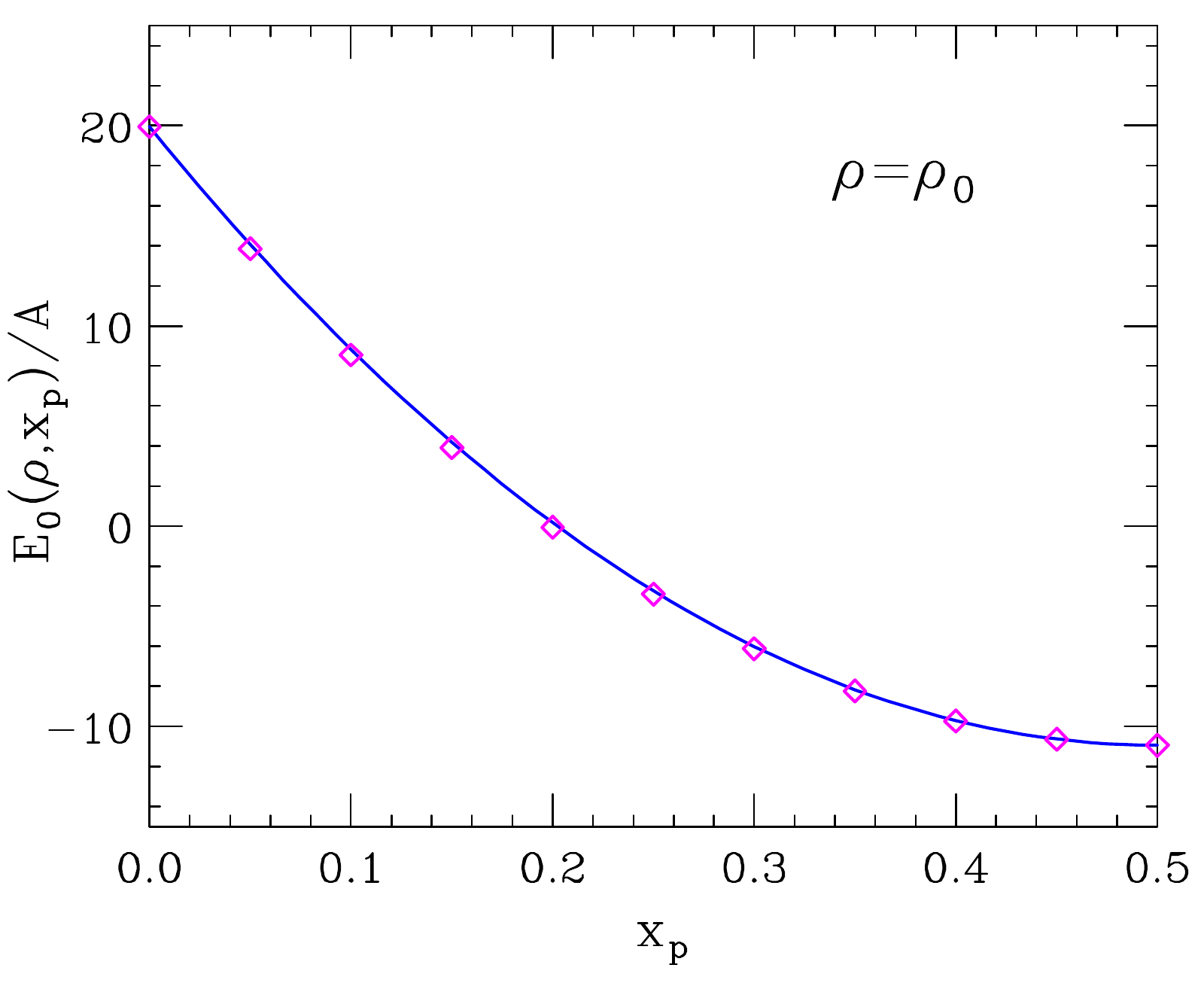}
\caption{Ground-state energy per nucleon of nuclear matter at baryon density $\rho=\rho_0$ and proton fraction 
$0 \leq x_p \leq 0.5$. The diamonds represent the results obtained using Eqs.~\eqref{eq:E_HF}-\eqref{slater} and the 
CBF effective interaction, while the solid line corresponds to the quadratic approximation of Eq.~\eqref{quadratic}. \label{x_dep}}
\end{figure}

As a final note, it has to be pointed out that our approach, allowing a straightforward calculation of the ground-state 
energy of nuclear matter as a function of both baryon density and neutron excess, is ideally suited to test the 
validity of the approximation of Eq.~\eqref{quadratic}. The results of Fig.~\ref{x_dep} clearly show that the quadratic 
approximation describes the $x_p$-dependence of the ground state energy at $\rho=\rho_0$, obtained from Eqs.\eqref{eq:E_HF}-\eqref{slater},
to remarkable accuracy. The deviation of the diamonds from the solid line turns out to be less than 3\% over the whole $x_p$ range.

\subsection{Pressure}

The pressure of nuclear matter, which plays a critical role in determining mass and radius of the equilibrium configurations 
of neutron stars, is simply related to the the ground-state energy through
\begin{align}
P = - \left(  \frac{\partial E_0}{\partial V} \right)_A =  \rho^2 \frac{ \partial (E_0/A) }{ \partial \rho } \ , 
\label{def:P}
\end{align}
where the derivative is taken keeping the number of nucleons constant.

\begin{figure}[bth]
\includegraphics[width=8.0cm]{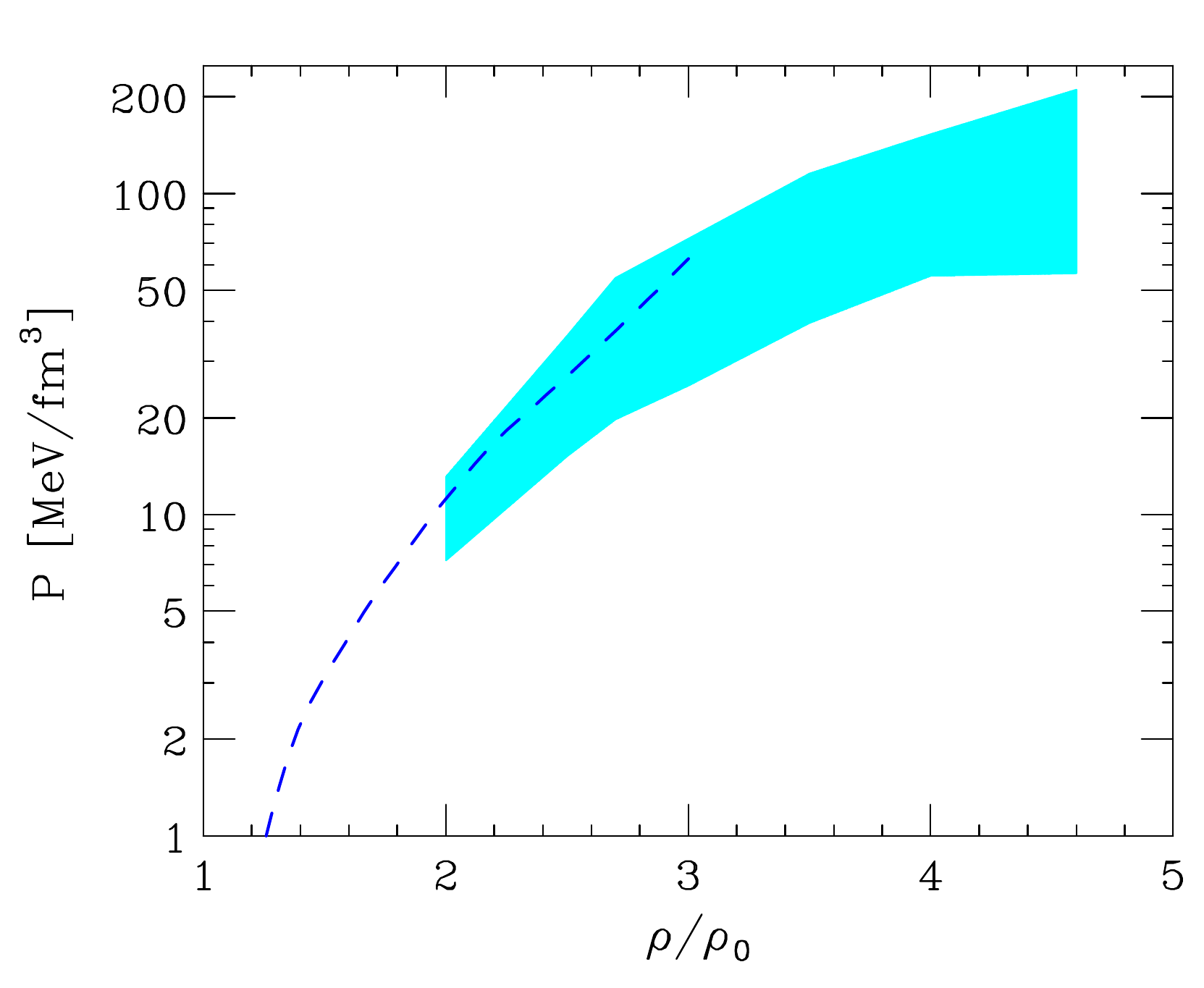}
\caption{ The dashed line illustrates the density dependence of the pressure of SNM obtained from the approach 
described in this paper. The shaded area corresponds to the region consistent with the experimental flow data reported 
in Ref.~\cite{danielewicz}.\label{P:SNM}}
\end{figure}

The dashed line of Fig.~\ref{P:SNM} illustrates the density dependence of the pressure of SNM obtained from our approach. For comparison, 
the shaded area shows the region consistent with the experimental flow data discussed in Ref.~\cite{danielewicz}, providing a constraint on  
$P(\rho)$ at  $\rho \geq 2 \rho_0$. It is apparent that, while being within the allowed boundary at  $2 \rho_0 \leq \rho \leq 3 \rho_0$, the calculated pressure
exhibits a slope suggesting that a discrepancy may occur at higher density. However, it has to be kept in mind that, being based on 
a non relativistic formalism, 
our approach is bound to predict a violation of causality, signalled by a value of the speed of sound in matter, defined as 
\begin{align}
v_s = \sqrt{ \frac{\partial P}{\partial (E_0/V) } }  \ , 
\end{align}
exceeding the speed of light in the high-density  limit. 

At equilibrium density, $v_s$ is trivially related to the compressibility 
modulus 
\begin{align}
\label{def:K0}
K_0 = \frac{1}{9} \left(  \frac{\partial P}{\partial \rho} \right)_{\rho = \rho_0}\ ,
\end{align}
which can be determined from measurements of the compressional modes in nuclei.
Using the SNM results reported in this article, we obtain the value $K_0~\approx~200$~MeV,  to be compared to the results of the analyses of Refs.~\cite{Shlomo2006,Colò2008}, 
yielding $K_0 = 240 \pm 20$ MeV.

\subsection{Single-particle spectrum and effective mass}

The conceptual framework for the identification of single-particle properties in interacting many-body systems is laid down in Landau's theory of Fermi 
liquids~\cite{baym-pethick}, based on the assumption that there is a one-to-one correspondence between the elementary excitations of a Fermi liquid, dubbed quasiparticles, 
and those of the noninteracting Fermi gas.

The energy of a quasiparticle of type $\lambda$ on the Fermi surface can be obtained by adding a particle of momentum $k=k_{F,\lambda}$ to
the system, without altering its volume. In the $A_\lambda = x_\lambda A \to \infty$ limit, this process leads to the expression 
\begin{align}
\label{HVH}
e_\lambda(k_{F,\lambda})  & = \left( \frac{ \partial E_0 }{ \partial A_\lambda } \right)_{V,A_{\mu \neq \lambda}} \\ \nonumber & = 
\left\{ \frac{ \partial  [ \rho (E_0/A) ] }{ \partial \rho_\lambda } \right\}_{V,\rho_{\mu \neq \lambda} } \ .
\end{align}
Note that the above equation, establishing a relation between the Fermi energy and the ground-state energy, is a straightforward generalization 
of the Hugenholtz-Van Hove (HVH) theorem~\cite{HVH}\textemdash one of the few exact results of the theory of interacting many-body systems\textemdash
to the multicomponent case.

The single-particle spectrum at fixed $\rho$, $e_\lambda(k)$, can be obtained following a process described 
by the authors of Ref.~\cite{friedpan}. Within this scheme, the energy of a quasiparticle (quasihole)
of momentum $k>k_{F,\lambda}$ ($k<k_{F,\lambda}$) is obtained moving a small fraction $\epsilon_\lambda$ of particles from a thin spherical shell at $k_{F,\lambda}$ ($k$) 
in momentum space to a thin spherical shell at $k$ ($k_{F,\lambda}$). Up to terms linear in $\epsilon_\lambda$, the resulting expression is
\begin{align}
e_\lambda(k) = e(k_{F,\lambda}) \pm \frac{1}{\epsilon_\lambda} \left[ \frac{ E(\epsilon_\lambda, k) }{A} - \frac{ E_0}{A} \right] \ ,
\end{align}
where the plus (minus) sign applies to the case $k>k_{F,\lambda}$ ($k<k_{F,\lambda}$). In the above equation, $E_0/A$ is the ground state 
energy per nucleon, while $E(\epsilon_\lambda, k)/A$ is the energy obtained modifying the Fermi gas density matrix according to 
\begin{align}
\label{newslater}
\ell(k_{F,\lambda} r) \to \ell(k_{F,\lambda} r) \pm \epsilon^\lambda \Big[ \frac{\sin(kr)}{kr} -  \frac{\sin(k_{F,\lambda} r)}{k_{F,\lambda} r} \Big] \ ,
\end{align}
where, once again,  the plus (minus) sign corresponds to $k~>~k_{F,\lambda}$ ($k<k_{F,\lambda}$). 

The above procedure, originally developed within the context of the variational FHNC/SOC approach, can be employed just as well to carry out perturbative calculations. 
At first order in the effective interaction, it reduces to using the modified density matrix of Eq.~\eqref{newslater} in Eq.~\eqref{eq:E_HF}, which 
 in turn  leads  to recover the expression of the single-particle energy in Hartree-Fock approximation 
\begin{align}
\label{eq:spec_hf}
e^{HF}_\lambda(k) =\frac{k^2}{2m} + \rho\sum_\mu & x_\mu \int d^3 r \Big[ v^\text{eff,d}_{\lambda\mu}({\bf r}) \\*
\nonumber
&  - v^\text{eff,e}_{\lambda\mu}({\bf r}) j_0(kr)  \ell(k_{F,\mu} r) \Big] \ , 
\end{align}
with $j_0(x) = \sin x/x$.

\begin{figure}[h!]
\includegraphics[width=8.0cm]{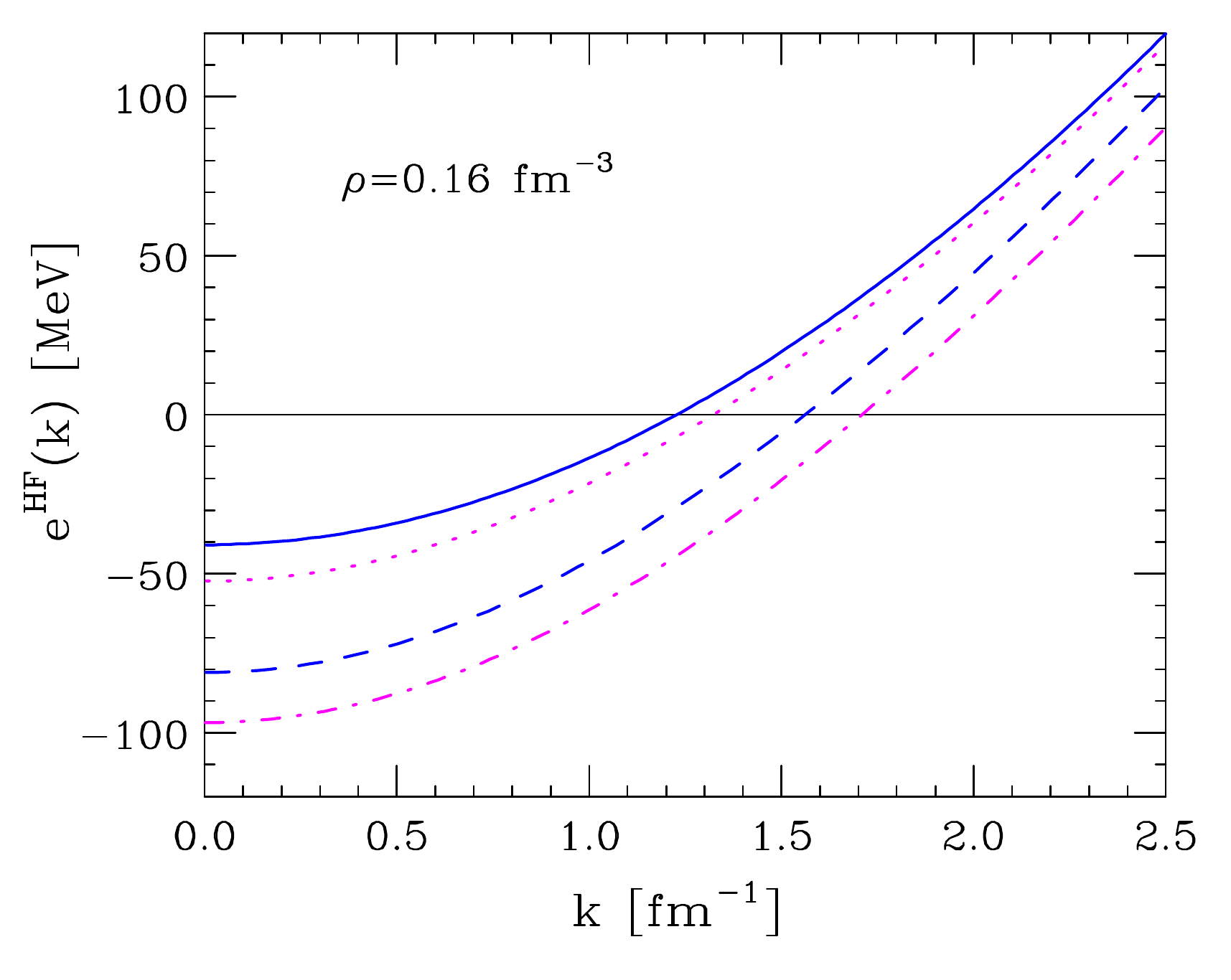}
\caption{Momentum dependence of the single-nucleon energies, evaluated at $\rho=\rho_0$ within the Hartee-Fock approximation of Eq.~\eqref{eq:spec_hf}. The solid and 
dashed lines correspond to PNM and SNM, respectively, whereas the dot-dash and dotted lines represent the proton and neutron spectra 
in matter with proton fraction $x_p = 0.1$. \label{e_k}}
\end{figure}

Figure~\ref{e_k} shows the momentum dependence of the Hartee-Fock spectra of protons and neutrons in nuclear matter, evaluated at $\rho=\rho_0$ and
$x_p =$ 0 (PNM), 0.1 and 0.5 (SNM). Note that the proton spectrum at $x_p=$0.1 is appreciably below the one corresponding to SNM. This feature, implying that 
neutron excess makes the mean field felt by a proton more attractive, is likely to be ascribed to the non-central component of the nuclear interaction. 
 
\begin{figure}[h]
\includegraphics[width=8.0cm]{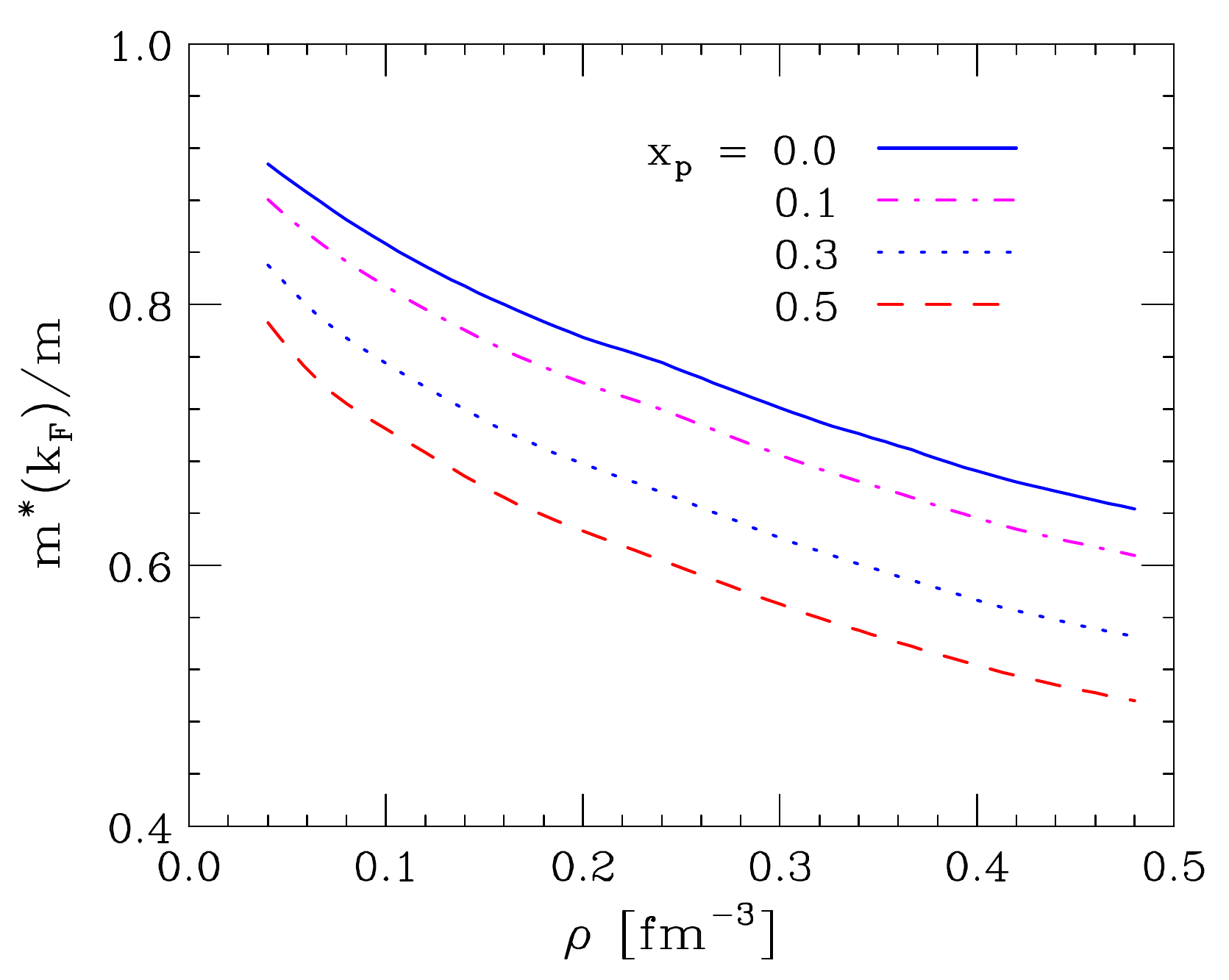}
\caption{Density dependence of the ratio $m^\star(k_F)/m$ for neutrons, computed using the Hartree-Fock spectra of  Eq.~\eqref{eq:spec_hf}. 
The solid and dashed lines correspond to PNM and SNM, respectively, while the dot-dash and dotted lines represent the results \label{mstar}
obtained setting $x_p=0.1$ and 0.3, respectively.}
\end{figure}

\begin{figure}[t!]
\includegraphics[width=8.0cm]{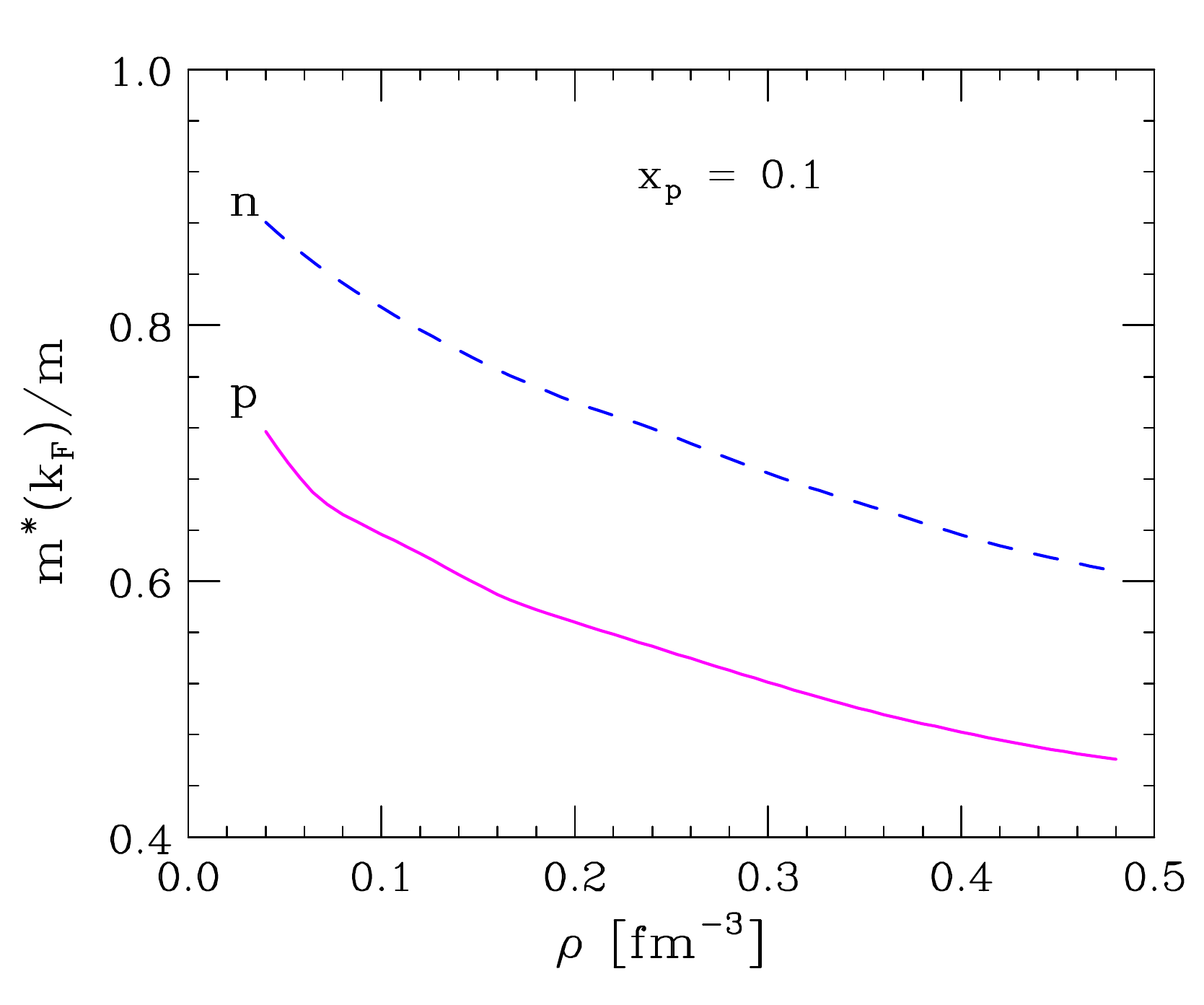}
\caption{Density dependence of the proton (p) and neutron (n) effective mass at the Fermi surface, computed setting the proton fracion to  $x_p=0.1$. 
 \label{mstar_np}}
\end{figure}

The single-particle energy is often paramertized in terms of the effective mass, defined by the equation
\begin{align}
\frac{1}{m^\star_\lambda(k)} = \frac{1}{m} \frac{d e_\lambda(k)}{d k} \ .
\label{mstar:def}
\end{align}
The density dependence of the neutron effective mass at $k=k_F$, obtained from the Hartree-Fock spectra of  Eq.~\eqref{eq:spec_hf}, is illustrated in Fig.~\ref{mstar} for different values of the proton fraction. The solid and dot-dash lines correspond 
to PNM and SNM, while the dotted and dot-dash lines have been obtained setting $x_p =$ 0.1 and 0.3, respectively. The difference between the proton and 
neutron effective masses in non-isospin-symmetric matter,  is illustrated in Fig.~\ref{mstar_np}, corresponding to proton fraction $x_p=0.1$.

\begin{center}
\begin{figure}[h!]
\includegraphics[width=7.cm]{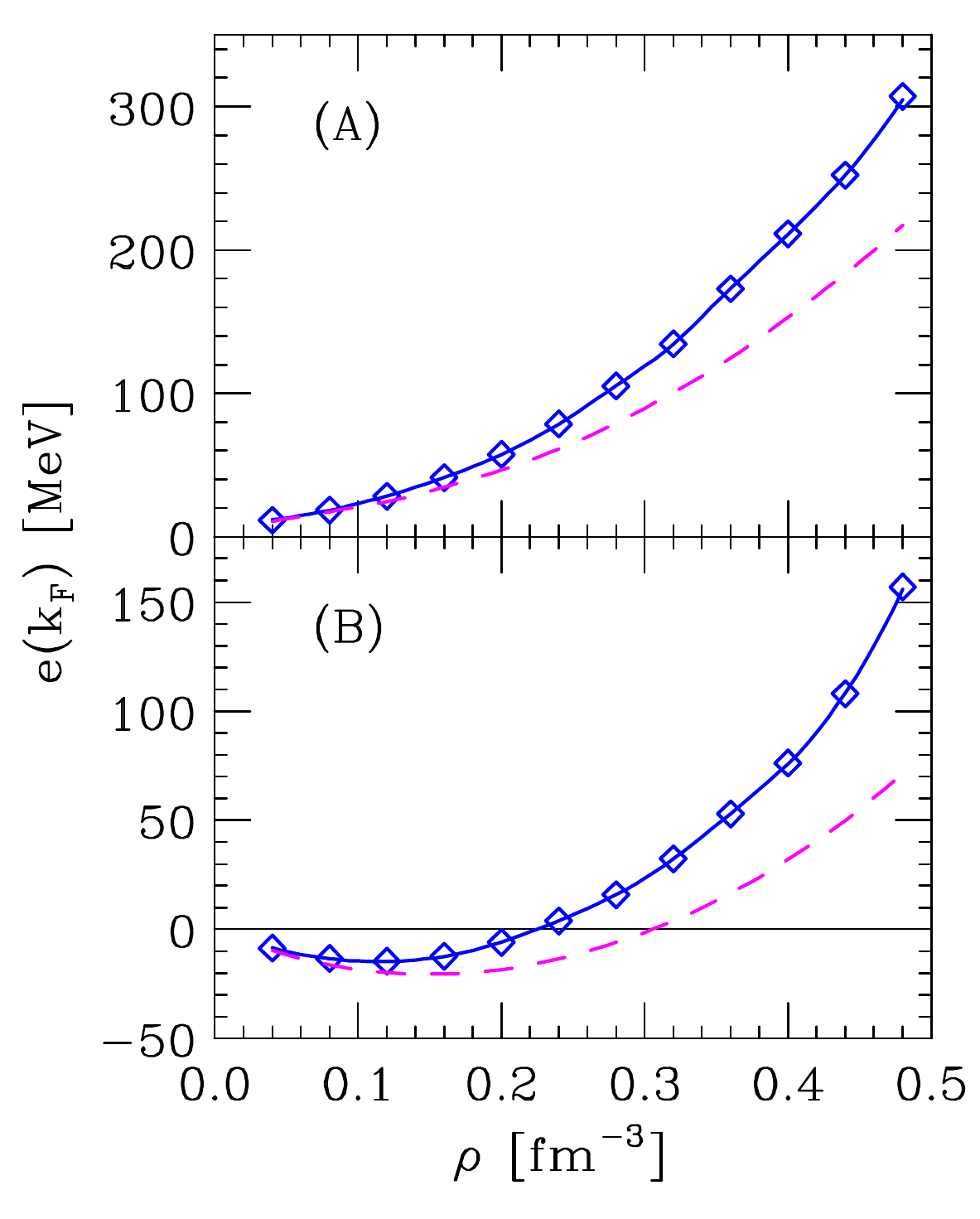}
\caption{Neutron energy at $k=k_F$ in PNM (A) and SNM (B). The solid lines have been obtained 
by differentiating the ground-state energy per nucleon according to Eq.~\eqref{HVH}. Dashed lines and diamonds correspond 
to the results of calculations performed within the Hartree-Fock approximation of Eq.~\eqref{eq:spec_hf} and including the 
rerrangement term according to Eq.~\eqref{eq:spec}, respectively. \label{fig:spec_hf}}
\end{figure} 
\end{center}

For density-independent interactions, the Hartree-Fock spectrum \eqref{eq:spec_hf} and the ground-state energy per nucleon of Eq.~\eqref{eq:E_HF} fulfill 
the requirement dictated by the HVH theorem by construction. On the other hand, it has long been recognized that large deviations from Eq.~\eqref{HVH} occur when the potential depends on $\rho$, as in 
the case of both the G-matrix~\cite{PhysRev.118.1438} and the CBF effective interaction. In order to restore consistency with the HVH theorem, the Hartee-Fock result must be corrected, by adding  a  
{\em rearrangement} term involving the derivative of $v^{\rm eff}$ with respect to the density $\rho_\lambda$. The resulting expression is
\begin{align}
 \label{eq:spec}
  \nonumber
e_\lambda(k_{F,\lambda}) & = e^{HF}_\lambda(k_{F,\lambda})  +\frac{1}{2} \sum_{\mu\nu} \rho_\mu \rho_\nu 
\int d^3 r \left[\, \left( \frac{\partial v^\text{eff,d}_{\mu\nu}({\bf r})}{\partial \rho_\lambda} \right) \right.\\*
& - \left. \left(  \frac{\partial v^\text{eff,e}_{\mu\nu}({\bf r})}{\partial \rho_\lambda} \right) \ell(k_{F,\mu} r)  \ell(k_{F,\nu} r)  \, \right] \ .
\end{align}
Figure~\ref{fig:spec_hf} shows the energy of a neutron carrying momentum $k=k_F$ in PNM (A) and SNM (B), computed using  Eqs.~\eqref{eq:spec_hf} (dashed lines) and \eqref{eq:spec} (diamonds). For comparison, the results obtained 
by differentiation of $\rho E_0/A$, as prescribed by the HVH theorem, are represented by the solid lines. It clearly appears that the Hartree-Fock 
approximation is only consistent at subnuclear densities. However, \textcolor{red}{the} inclusion of the rearrangement term\textemdash whose size increases from
$\sim$5 Mev to $\sim$80 MeV in the density range $1 \lesssim \rho/\rho_0 \lesssim 3$\textemdash brings the Fermi energies into 
perfect agreement with the predictions of Eq.\eqref{HVH}.

The impact of the rearrangement correction on the momentum dependence of the single-nucleon energy in nuclear matter has been thoroughly 
discussed in the context of G-matrix perturbation theory~\cite{PhysRev.112.906,PhysRev.118.1438}. The authors of Ref.~\cite{HVH2} argued that 
in the vicinity of the Fermi surface, that is, at $k \approx k_{F,\lambda}$, the spectrum can be obtained from the simple approximate expression 
[compare to Eq.\eqref{eq:spec}]
\begin{align}
 \label{eq:alfredo}
  \nonumber
e_\lambda(k) & \approx e^{HF}_\lambda(k)  +\frac{1}{2} \sum_{\mu\nu} \rho_\mu \rho_\nu 
\int d^3 r \left[\, \left( \frac{\partial v^\text{eff,d}_{\mu\nu}({\bf r})}{\partial \rho_\lambda} \right) \right.\\*
& - \left. \left(  \frac{\partial v^\text{eff,e}_{\mu\nu}({\bf r})}{\partial \rho_\lambda} \right) \ell(k_{F,\mu} r)  \ell(k_{F,\nu} r)  \, \right] \ .
\end{align}
From the definition of Eq.\eqref{mstar:def},  it follows  that according to the above prescription the ratio $m^\star(k_{F,\lambda})/m$\textemdash which plays a
driving role in a number of processes of astrophysical interest\textemdash is not affected by the rearrangement term. 

\section{Summary and outlook}
\label{summary}

An improved version of the effective interaction derived in Refs.~\cite{PhysRevLett.99.232501,Lovato:2012ux,Lovato:2013dva}\textemdash obtained from 
a microscopic nuclear Hamiltonian 
using the CBF formalism and the cluster expansion technique\textemdash has been employed to 
perform perturbative calculations of several properties of nuclear matter at arbitrary neutron excess.

The well behaved CBF effective interaction embodies all the distinctive features of the bare interaction, as well as the
screening effects associated with the repulsive core. In addition, unlike the effective interactions specifically designed to 
reproduce the bulk properties of nuclear matter\textemdash see, e.g., Refs.\cite{skyrme1,skyrme2}\textemdash it can be used to describe 
nucleon-nucleon scattering in the nuclear medium, whose understanding is needed for the description of 
non-equilibrium properties relevant to astrophysical processes.

It has to be pointed out that the CBF effective interaction is {\em not} defined in operator form, but only in terms of
its expectation value in the Fermi gas ground state. However, the assumption that perturbative calculations
involving matrix elements of $v_{ij}^{\rm eff}$ between Fermi gas states provide accurate estimates of nuclear matter properties other than the ground-state energy 
is strongly supported by the results of recent studies of the Fermi hard-sphere system~\cite{mecca_1,mecca_2}.

While admittedly failing to precisely reproduce the empirical value of the ground-state energy of SNM\textemdash mainly because of deficiencies 
of the bare Hamiltonian\textemdash our approach
predicts the correct equilibrium density, as well as reasonable values of both the symmetry energy and the compressibility. 
Moreover, it is perfectly suited to describe spin-polarized matter.

In the future, the accuracy of the CBF effective interaction approach may be improved using the coordinate-space nuclear Hamiltonians recently derived 
within chiral perturbation theory~\cite{Gezerlis:2013,gezerlis:2014,piarulli:2015,piarulli:2016,tews:2016,Lynn:2016,logoteta:2016}.   However, 
we believe that, 
in view of the broad range of possible astrophysical applications\textemdash most notably studies of neutron star structure and dynamics and supernova explosions\textemdash  
the availability of a theoretical framework allowing for a 
consistent treatment of a broad range of nuclear matter properties within a unified model of nuclear dynamics will prove critically important. In this context, 
a $\sim 15\%$ error in the ground-state expectation value of the potential energy of SNM at saturation density appears
to be an acceptable price to pay.  

As a final remark, it has to be pointed out that, as long as thermal effects do not lead to modifications of the underlying strong interaction dynamics, the formalism described in this article
can be readily generalized to treat nuclear matter at nonzero temperature, by replacing the $T=0$ Fermi distribution appearing in the right-hand side of Eq.~\eqref{slater}
with the corresponding distribution at temperature $T>0$. 

Preliminary results of the extension of the CBF effective interaction approach to the treatment of hot nuclear 
matter\textemdash the details of which will be discussed elsewhere\textemdash have been employed by the authors of Ref.~\cite{Camelio:2017nka} to study the neutrino
luminosity and gravitational wave emission of proto-neutron stars during the Kelvin-Helmoltz evolutionary phase.  

\appendix
\section{Matrix elements of the effective interaction in spin-isospin space}
\label{matrices}

In this Appendix, we provide the explicit expressions of the quantities needed for the calculation of the 
matrix elements of the effective interaction in spin-isospin space. They can be conveniently rewritten in the 
form
\begin{align}
\nonumber
v^\text{eff,d}_{\lambda\mu}({\bf r}_{ij}) & = \sum_{p} v^{p}(r_{ij})  A^{p}_{\lambda \mu}(\cos \theta) \ , \\
\nonumber
v^\text{eff,e}_{\lambda\mu}({\bf r}_{ij}) & = \sum_{p} v^{p}(r_{ij})  B^{p}_{\lambda \mu}(\cos \theta)  \ , 
\end{align}
where 
\begin{align}
\nonumber
A^p(\cos \theta) = \langle \lambda\mu | O^{p}_{ij} | \lambda\mu\rangle \ , \\
\nonumber
B^p(\cos \theta) = \langle \lambda\mu | O^{p}_{ij} | \mu \lambda \rangle \ , 
\end{align}
$\cos \theta={\bf r}_{ij}/|{\bf r}_{ij}|$ and the operators $O^{p}_{ij}$, with $p=1,\ldots,6$, are given by Eqs.\eqref{av18:2} and \eqref{S12}.

The matrices $A^p(\cos \theta)$ and $B^p(\cos \theta)$ read 
\begin{align}
\nonumber
A^1 = \left(
\begin{array}{cccc}
1 & 1 & 1 & 1 \\
1 & 1 & 1 & 1 \\
1 & 1 & 1 & 1 \\
1 & 1 & 1 & 1 \\
\end{array}
\right) \  , \   
A^2 = \left(
\begin{array}{rrrr}
1 & 1 & -1 & -1 \\
1 & 1 & -1 & -1 \\
-1 & -1 & 1 & 1 \\
-1 & -1 & 1 & 1 \\
\end{array}
\right)  \ , 
\end{align}
\begin{align}
\nonumber
A^3  = \left(
\begin{array}{rrrr}
1 & -1 & 1 & -1 \\
-1 & 1 & -1 & 1 \\
1 & -1 & 1 & -1 \\
-1 & 1 & -1 & 1 \\
\end{array}
\right) \ , 
\end{align}
\begin{align}
\nonumber
A^4 = \left(
\begin{array}{rrrr}
1 & -1 & -1 & 1 \\
-1 & 1 & 1 & -1 \\
 -1 & 1 & 1 & -1 \\
 1 & -1 & -1 & 1 \\ 
\end{array}
\right) \ , 
\end{align}
\begin{align}
\nonumber
A^5 = A^2 \ ( 3 \cos \theta^2 - 1 ) \ , 
\end{align}
\begin{align}
\nonumber
A^6 = A^4 \ ( 3 \cos \theta^2 - 1 ) \ , 
\end{align}
and
\begin{align}
\nonumber
B^1 = \left(
\begin{array}{cccc}
1 & 0 & 0 & 0 \\
0 & 1 & 0 & 0 \\
0 & 0 & 1 & 0 \\
0 & 0&  0 & 1 \\
\end{array}
\right) \  , \   
B^2 = \left(
\begin{array}{rrrr}
1 & 0 & 2 & 0 \\
0 & 1 & 0 & 2 \\
2 & 0 & 1 & 0 \\
0 & 2 & 0 & 1 \\
\end{array}
\right)  \ , 
\end{align}
\begin{align}
\nonumber
B^3 = \left(
\begin{array}{cccc}
1 & 2 & 0 & 0 \\
2 & 1 & 0 & 0 \\
0 & 0 & 1 & 2 \\
0 & 0&  2 & 1 \\
\end{array}
\right) \  , \   
B^4 = \left(
\begin{array}{rrrr}
1 & 2 & 2 & 4 \\
2 & 1 & 4 & 2 \\
2 & 4 & 1 & 2 \\
4 & 2 & 2 & 1 \\
\end{array}
\right)  \ , 
\end{align}
\begin{align}
\nonumber
B^5  = \left(
\begin{array}{rrrr}
1 & -1 & 0 & 0 \\
-1 & 1 & 0 & 0 \\
0 & 0 & 1 & -1 \\
0 & 0 & -1 & 1 \\
\end{array}
\right)  ( 3 \cos \theta^2 - 1 ) \ , 
\end{align}
\begin{align}
\nonumber
B^6  = \left(
\begin{array}{rrrr}
1 & -1 & 2 & -2 \\
-1 & 1 & -2 & 2 \\
2 & -2 & 1 & -1 \\
-2 & 2 & -1 & 1 \\
\end{array}
\right)  ( 3 \cos \theta^2 - 1 ) \ . 
\end{align}


\acknowledgments
The authors are deeply indebted to Giovanni Camelio, for countless stimulating discussions on topics related
to the subject of this Article. 
The work of OB is supported by INFN under grant MANYBODY.
The work of AL  is supported by the U.S.~Department of Energy, Office
of Science, Office of Nuclear Physics, under contracts DE-AC02-06CH11357.

%


\end{document}